\documentclass[]{aa} 
\usepackage{rotating}

\usepackage{graphicx}

\def\cmM2{\rm cm$^{-2}$}
\def\etal{et~al.~}
\def\leda{LEDA170194}
\def\075{IGR~J07565-4139}
\def\120{IGR~J12026-5349}
\def\j104{IGR~J10404-4625}
\def\ngc{NGC~788}
\def\eso{ESO~103-G35} 
\def\ic{IC~4518A}
\def\flux{\rm erg\ cm^{-2}\ s^{-1}}
\def\chandra{{\it Chandra}}
\def\integral{{\it INTEGRAL}}
\def\xmm{{\it XMM-Newton}}
\def\sax{{\it Beppo-SAX}}
\def\asca{{\it ASCA}}
\def\swift{{\it Swift/\rm BAT}}
\begin{document}

\title{An X-ray  view of absorbed \integral\ AGN}
\author{A. De Rosa \inst{1}, L. Bassani \inst{2}, P. Ubertini \inst{1}, F. Panessa \inst{1}, A. Malizia \inst{2},  A. J. Dean \inst{3}, \and R. Walter \inst{4}
     	 }
	
\institute
{INAF/IASF, via del Fosso del Cavaliere 100, I-00133 Rome, Italy
\and INAF/IASF, via Gobetti 101,I-40129 Bologna, Italy
\and School of Physics and Astronomy, University of Southampton
	Highfield, Southampton, United Kingdom,
\and INTEGRAL Science Data Centre, Chemin d'Ecogia 16, CH-1290 VERSOIX, Switzerland
}

\offprints{alessandra.derosa@iasf-roma.inaf.it}
\date{Received   /Accepted   }

\authorrunning{De Rosa et al.}

\abstract
{}
{We  present a 0.2--200 keV broad-band study of absorbed AGN observed with \textit{INTEGRAL}, \textit{XMM-Newton}, \textit{Chandra} and \textit{ASCA} to investigate the continuum shape and the absorbing/reflecting medium properties.}
{The sources are selected in the \integral\ AGN sample to have a 20--100 keV flux below 8$\times$10$^{-11}$ $\flux$ (5 mCrab), and are characterized by a 2--10 keV flux in the range (0.8--10)$\times$10$^{-11}$ $\flux$. The good statistics allow us a detailed study of the intrinsic and reflected continuum components. In particular, the analysis performed on the combined broad-band spectra allow us to investigate the presence of Compton reflection features and high energy cut-off in these objects.}
{The column density of the absorbing gas establishes the Compton thin nature for three sources in which a measure of the absorption was still missing. The Compton thin nature of all the sources in this small sample is also confirmed by the diagnostic ratios F$_{\rm x}$/F$\left[\rm O  III\right] $.  The Compton reflection components we measure, reflection continuum and iron line,  are not immediately compatible with a scenario in which the absorbing and
reflecting media are one and the same, i.e. the obscuring torus. A possible solution is that the absorption is more effective than reflection, e.g. under the hypothesis that the absorbing/reflecting medium is not uniform, like a clumpy torus, or that the source is observed through a torus with a very shallow opening angle. 
The high energy cut-off (a lower limit in two cases) is found in all sources of
our sample and the range of values is in good agreement with that found in type 1 Seyfert galaxies.
At lower energies there is clear evidence of a soft component (reproduced with a thermal and/or scattering model), in six objects.}
{}


\maketitle

\section{Introduction}

The broad-band properties of Seyfert 2 AGN have been extensively studied in the past years mainly using \sax\ data (\cite{risaliti02}).
The continuum shape in these objects can be reproduced by a steep power-law with photon index $\Gamma \sim$ 2. 
This model matches well a Comptonization scenario in which seed photons from a cold gas (the accretion disc) are up-scattered
in a hot corona of relativistic electrons (\cite{HM91}, \cite{HMG97}).
A photoelectric cut-off is present at X-ray energies and its value depends on the absorbing column density of the source. For N$_H$ in the $10^{22}-10^{23}$ \cmM2 range
the cut-off is below 2 keV, for N$_H$ in  the $10^{23}-10^{24}$ \cmM2 range  the cut-off is in 3--10 keV band and for N$_H >\sigma_{T}^{-1}=1.5\times 10^{24}$ \cmM2 
(i.e. the Compton thick regime with  $\sigma_{T}^{-1}$ being the inverse of the Thomson cross section), 
the continuum can be observed only through its reflected component above 10 keV.

The presence of a high energy cut-off is still an open issue. Risaliti (2002)  found that the high energy cut-off 
is present only in $\sim$ 30 
per cent of a sample of 20 bright Sy 2 observed with \sax. Above 10 keV the presence of a Compton reflection component has often been observed, however is its origin still unclear as well as if the reflecting/absorbing medium are one and the same (the putative ''molecular torus''). 
A cold iron line at 6.4 keV is often associated to this continuum Compton reflection hump.

The IBIS gamma-ray imager on board \integral\ is surveying 
the whole sky above 20 keV with a mCrab ($\sim $10$^{-11}$ erg cm$^{-2}$ s$^{-1}$) sensitivity in well 
exposed regions. With its angular resolution
of 12 arcmin and the point source location accuracy of 1-2 arcmin for
moderately bright sources (Ubertini et al. 2003) \integral\ has already
detected a number of AGN,
a large fraction  of which were previously unknown as high energy emitters 
(\cite{bird06}, \cite{bassani06a}, \cite{bird07}).

In this paper we present a deep broad-band spectral analysis of seven absorbed AGN detected by
\integral. 
The sample is extracted from the Bassani et. al (2006a) survey, updated to include a
number of optical classifications obtained afterwards (see \cite{bassani06b}). The sample includes
Seyfert 2 galaxies having a 20--100 keV flux below 5 mCrab and with X-ray data
available at the time that this work started; from the sample we have excluded sources
already studied by \sax\ except for one object (\eso) which was retained in
order to provide a direct comparison between this work
and previous studies using  \sax\ data  and therefore an \textit{a posteriori} check of
our analysis procedure
which is affected by the limitation of using non simultaneous X and soft gamma-rays
data. Overall we can conclude that the sample used in this study is representative of the population of type 2 AGN detected by \integral\ above 10 keV; further broad band X/gamma-ray analysis on a complete sample of type 2 AGN selected in the 20-40 keV band is on going and will be the objective of a future paper\footnote{6 other type 2 AGN from Bassani et al. survey are below the assumed threshold flux:
3 (IGR J12415+5750, IGR J20286+2544 and IGR J17513-2011) had no X-ray available
data when this
analysis started while 3 (NGC 1068, Mkn 3 and NGC 6300) had BeppoSAX observational
results published but all are peculiar objects  the first two being  Compton thick
AGN and the third one the prototype of "changing look" Seyfert 2 ( i.e. a source
that   goes from thick to thin and viceversa).}.
Four sources (IGR J12391-1610=\leda\ hereafter, \075, \120 and \j104) have been discovered first in soft gamma-ray band and  immediately after observed with \chandra\ (\leda, \075 and \120) and \xmm\ (\j104). Two objects (\eso\ and \ic) are known AGN (although \ic\ was never reported at X-ray energies before) and follow up observations with \xmm\ have been used in our broad-band analysis.
For one object (\ngc) archival \asca\ data have been used in this work. 
Using X-ray measurements in conjunction with \integral\ data allow us to build a spectrum in three decades of energy, and to distinguish between the various spectral components characterizing the broad-band spectra of absorbed AGN. This in turn will allow us, in particular, to study the absorption/reflection medium properties of the sample. These good quality spectra also permit a detailed investigation of the intrinsic continuum slope including the cut-off if present. The broad--band coverage available is a powerful and unique tool to address all these issues, making possible a simultaneous measurement of the intrinsic continuum (both photon index and high energy cut-off) and of the reflected continuum together with the Fe line.

Note that a number of sources in the sample have data already published in the
literature both in the X-ray band
(Sazonov et al. 2005, Shinozaki et al. 2006) and in soft gamma-ray range (Beckmann et
al. 2006, Molina et al. 2006).
However, most of these works did not provide a detailed analysis of the broad band
spectra of the sources in the sample nor attempted  to put constraints on the high
energy cut-off and reflection bump. The only exception
is \eso\ analysed by our team using \sax\ and \integral\ data; here we use a
recent \xmm\ observation
of the source in conjunction with an updated \integral\ spectrum  in order to have in the
sample a well studied object to use as a source of reference.

In Sect. \ref{data} we present the data of our small sample while the models used to fit the spectra are described in Sect.
\ref{models}.
Results and discussion are presented in Sect. \ref{RD} and draw our conclusions. 
The single cases are discussed in more details in the appendix.

\section{Observations and data reduction}
\label{data} 
\subsection{\integral\ data}

The AGN  discussed here have been observed with 
IBIS (Ubertini et al. 2003), the imager on board \emph{INTEGRAL} (Winkler et al. 2003). 
This coded mask instrument is made by the combination
of two detector layers: ISGRI (Lebrun et al. 2003), an upper CdTe
layer sensitive in the range between 15 keV and 1 MeV, and PICsIT (Di Cocco et
al. 2003) a bottom CsI layer sensitive in the range 200 keV to 8 MeV.
In the present paper, we refer to data collected by the first layer
only, since the sources are too weak above $\sim$ 100 keV for detection by PICsIT.

For each source,  \integral\ data from  several pointings performed between
revolution 12 and 429 were added together in the following way to provide sufficient
statistics.
First \emph{ISGRI} images for each available pointing were generated in 13 energy
bands using the ISDC offline scientific analysis software OSA
version 5.1. Then, count rates at each source position were extracted
from individual images in order to provide light curves
in the various energy bands sampled; since the light curves did not show any sign of variability or flaring activity, average fluxes were
then extracted in each band and combined to produce the source spectrum. 
A detailed description of the
source extraction criteria can be found in Bird et al. (2006, 2007).
We have tested that this method of spectral extraction is
reliable by comparing the Crab spectrum obtained in this way with the one extracted
using the standard spectral analysis.

In Table \ref{jurnal} we report the details of the IBIS/ISGRI observations;
best fit positions, exposures and count rates in the range 20--100 keV. The positional uncertainty for sources of this intensity is around $\sim$3 arcmin (Bird et al. 2006). \\

\subsection{\chandra\ data} 

\chandra\ observed \leda\ on July 25, 2005, and \120 and \075 on June 16, 2005, with ACIS-I. Details on the observations are given
in Table \ref{jurnal}. 
Data were reduced using CIAO v3.2, following standard procedures. 
It is important to note here that although Sazonov et al. (2005) used \chandra\
spectra in conjunction with \integral\
data, they did not attempt to estimate cut-off energy and reflection bump in their
analysis probably due to the limited
amount of spectral information available above 10 keV (a single bin in the 17-60 keV
band). 
All \chandra\ sources suffer from pile-up. 
Using WebPIMMS\footnote{http://heasarc.gsfc.nasa.gov/Tools/w3pimms.html} we estimated
the fraction of pile-up to be around 36 per cent for \075, 30 per cent for \120 and 
$>$20 per cent for \leda.
The effect of pile-up is clearly visible through an hardening
of the spectrum above 4.5-5 keV. 
Regarding the pile-up problem, we point out that we have used a different approach than
the one adopted by Sazonov et al. (2005): these authors removed the central pixel, i.e. excluding 80\% of the instrument PSF and consequently used data up to 5 keV; we have instead maintained
the central pixel and dealed with the spectra affected by pile-up through the PILE-UP model available in XSPEC which can correct the total spectrum for the pile-up effects, provided that this is less than 70\% (Davis 2001).

\subsection{\xmm\ data}

The \xmm\  EPIC observation of \eso, \ic\ and \j104\ have been also. Details on the observations are given in Table \ref{jurnal}. 
Data reduction was performed with version 7.0 of the SAS software,
employing the most updated calibration files available at the time of the data
reduction (2006 March).
Patterns 0 to 12 (4) were employed in the extraction of the MOS (pn)
scientific products. In the case of \ic\ pn and MOS spectra for the source were
extracted from a circular region of
30 arcsec radius while background spectra were taken 
from source-free circular regions of 10 arcsec radius.
In the case of \eso\ MOS (pn) spectra were
extracted from a circular region of
35 (60) arcsec radius around the source while background spectra were taken
from source-free circular regions of 10 (40) arcsec radius.
Finally in the case of \j104 MOS (pn) spectra were
extracted from a circular region of
40 (90) arcsec radius around the source while background spectra were extracted
from source-free circular regions of 20 (40) arcsec radius.
Spectra are rebinned to have at least 20 counts in each spectral
channel.
Spectral fit were performed simultaneously with MOS1, MOS2 and pn in
0.5--10 keV. 
Cross-calibration constant pn/MOS1 and MOS2/MOS1 were
left free to vary and always found in the range 0.97-1.04.
\xmm\ data of \eso\ have been already published in Shinozaki et al. (2006). Their analysis was restricted to 0.5--10 keV energy range and is consequently not conclusive about Compton reflection and high energy cut-off. In Appendix \ref{single sources} we will discuss also our results about absorption in comparison with those presented in Shinozaki et al. (2006).

\subsection{\asca\ data}
We finally analysed an archival \asca\ spectrum of \ngc.
The relevant spectra and associated files were downloaded from the TARTARUS archive\footnote{http://tartarus.gsfc.nasa.gov}.
We fitted the spectra of all instruments SIS0, SIS1, GIS2 and GIS3 simultaneously in the 0.6--10 keV band.
The cross-calibration constant between the different instruments were left
free during the fit and found in the range SIS1/SIS0=1.06$^{+0.11}_{-0.11}$, GIS2/SIS0=0.96$^{+0.10}_{-0.09}$, GIS3/SIS0=1.09$^{+0.10}_{-0.09}$.

\begin{table*}
\begin{flushleft}
\caption{Journal of the observations}
\begin{tabular}{lccccc}
\noalign{\hrule}
\noalign{\medskip}
\textbf{Name}      & \textbf{Mission} & \textbf{Date} & \textbf{Exp Time} & \textbf{RA, DEC (J2000)} & \textbf{Counts}\\
               &           & &  (s)  & (h m s, $^{\circ}$'") & (s$^{-1}$)\\
\hline 
\leda &  \chandra & 2005/07/25 &  3213  &  12 39 06.24,-16 11 56.4 & $^\star$0.063$\pm$0.005 \\
\leda & \integral & -  &  200000 &  12 39 11.04, -16 10 55.2 & $^{\circ}$0.76$\pm$0.09 \\
\hline 
\075 &  \chandra   & 2005/06/16 &  3216    &  07 56 15.36, -41 38 09.6 & $^\star$0.215$\pm$0.008\\
\075 & \integral & - & 968000 &   07 56 29.52, -41 38 31.2 & $^{\circ}$0.16$\pm$0.03 \\
\hline 
\120 & \chandra  & 2005/06/16 & 3220 & 12 02 36.72, -53 49 12.0 & $^\star$0.209$\pm$0.008\\
\120 & \integral & - &728000 & 12 02 50.16, -53 49 12.0 & $^{\circ}$0.47$\pm$0.05\\
\hline 
\ngc & \asca& 1999/01/16-18 & 39160 & 02 01 04.30, -06 52 48.0 & $^{\dagger}$0.024$\pm$0.001 \\
\ngc & \integral & - & 594000 & 02 01 03.36, -06 48 50.4 & $^{\circ}$0.85$\pm$0.07\\
\hline
\eso & \xmm&  2002/03/15 & 11987 &  18 38 08.30, -65 24 28.8 & $^\ddagger$0.507$\pm$0.007(1.71$\pm$0.01)\\
\eso & \integral & - &44000 & 18 38 46.80, -65 24 28.8 & $^{\circ}$0.84$\pm$0.13\\
\hline
\ic & \xmm & 2006/08/07 & 11460 & 14 57 44.32, -43 09 33.8 & $^\ddagger$0.061$\pm$0.03(0.218$\pm$0.007)\\
\ic & \integral & - &898000 &  14 57 38.40, -43 07 44.4 &  $^{\circ}$0.293$\pm$0.03\\
\hline
\j104 & \xmm & 2006/11/29 & 13530 &  10 40 22.32,  -46 25 25.6 & $^\ddagger$ 0434$\pm$0.003(1.40$\pm$0.01)\\
\j104 & \integral & - &626000 &   10 40 25.68,  -46 24 36.0 &  $^{\circ}$0.283$\pm$0.04\\
\hline
\end{tabular}
\label{jurnal}
\end{flushleft}
\small{$^\star$ In 1-7 keV. $^{\dagger}$ In 2-10 keV for SIS0. $^\ddagger$ in 0.5-10 keV
  for MOS(pn). $^{\circ}$ In 20-100 keV for IBIS/ISGRI.}
\end{table*}

\section{Spectral fitting}
\label{models}
 
The spectral analysis has been performed with XSPEC v11.3. Errors
correspond to the 90 per cent confidence level for one interesting parameter
($\Delta\chi^2$=2.7). We fitted simultaneously the soft and hard
X-ray spectra available with \integral\ and either \chandra, or \xmm\ or
\asca. 
The cross-calibration constant between \integral\ and the low energies instruments (C$_{INTEGRAL}$) was
fixed to 1.
Possible miscalibration  between \integral\ and \xmm, \chandra\ or \asca\  (i.e. C$_{INTEGRAL}\neq$ 1) can mimic or hide the presence of a Compton reflection component
above 10 keV. We choose to put this constant equal to one for all spectra because the fit we found with this value is good (all
reduced $\chi^2$ around  1). However, in Sect. \ref{RD} we discuss the implications on our fit if C$_{INTEGRAL}$ is left free to vary.

\begin{figure*}[!]
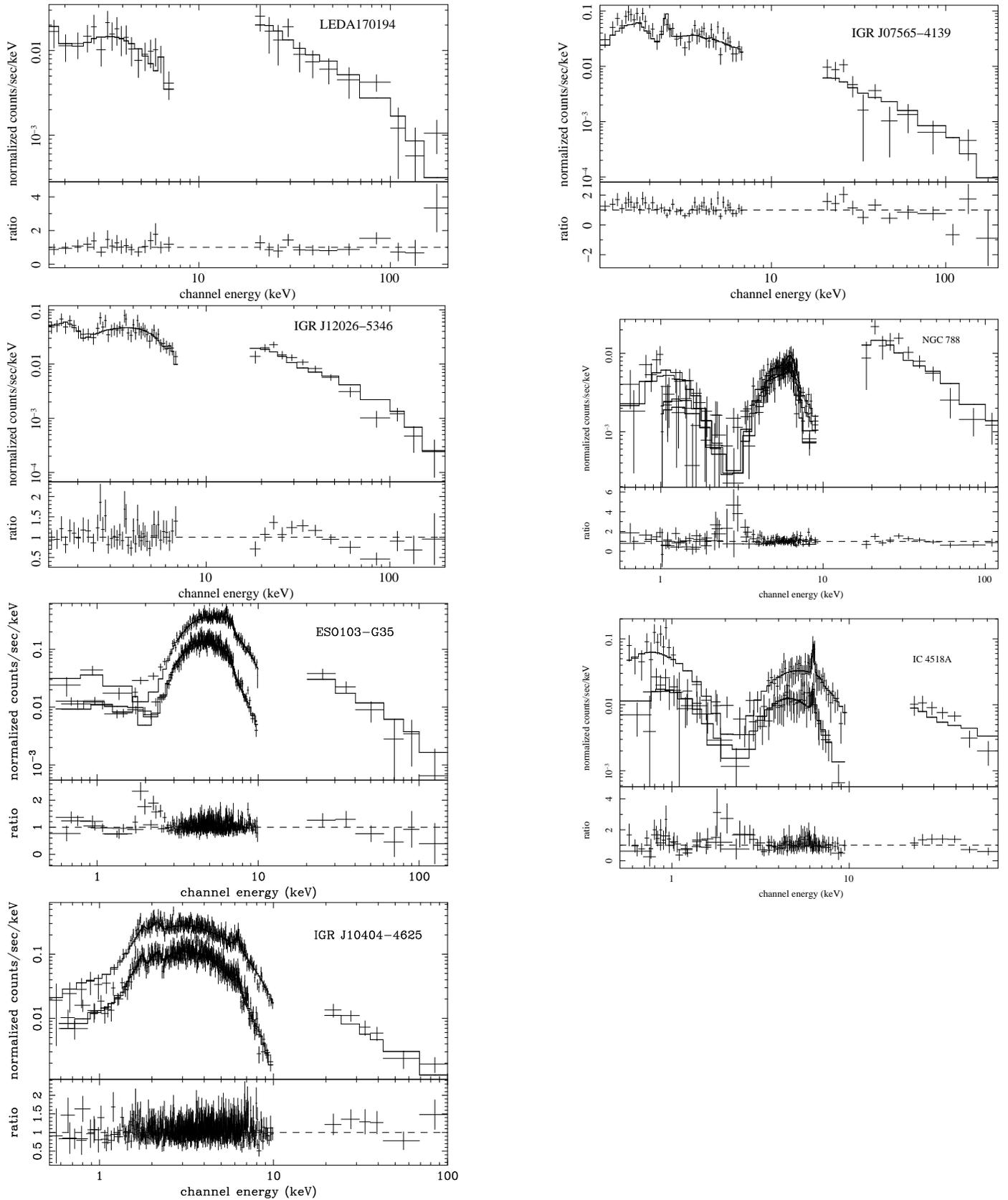

\includegraphics[width=0.3\linewidth,angle=-90]{ld_ra_leda_pil.ps}
\includegraphics[width=0.3\linewidth,angle=-90]{ld_ra_075_pil.ps}
\includegraphics[width=0.3\linewidth,angle=-90]{ld_ra_120_pil.ps}
\includegraphics[width=0.3\linewidth,angle=-90]{ld_ra_ngc.ps}
\includegraphics[width=0.3\linewidth,angle=-90]{ld_ra_eso.ps}
\includegraphics[width=0.3\linewidth,angle=-90]{ld_ra_ic.ps}
\includegraphics[width=0.3\linewidth,angle=-90]{ld_ra_104.ps}
\caption{Spectra and data/model ratio for all sources of our sample. The ratios are plotted using our model A: absorbed power-law+thermal emission+iron line.}
\label{ld ratio}
\end{figure*}

To reproduce the broad-band spectra of our sample, we employed
three different models. The first one (Model A) consists of
an intrinsic absorbed power-law, a thermal soft X-ray component and a gaussian emission (Fe)
line. 
The data and data/model A ratio for all source is shown in Figure \ref{ld ratio}.
In the second model (Model B), we added to the primary emission a high
energy cut-off; in this case the primary emission was modelled with 
an e-folded power-law (CUTOFFPL model in XSPEC). 
In the third model (Model C), we added to the previous continuum a Compton
reflection component;
in this model the primary emission is modelled with an e-folded
power-law plus reflection from an infinite cold slab.
In the Compton reflection component (PEXRAV model in XSPEC,
\cite{pexrav_ref}), the cosine of inclination angle of the
reflector was fixed to $0.95$. We assume elemental
abundances from Anders \& Grevesse (1989).
In the case of \leda, \075\ and \120 we choose to fix in all models the photon index
$\Gamma$=1.7, i. e. the average value observed in AGN, the energy of the Fe line  E$_{Fe}$=6.4 keV and the intrinsic width of the line $\sigma_{Fe}$=10 eV. These assumptions are needed because the low statistics of \chandra\ data did not allow us to constrain simultaneously all the parameters of the models, $\Gamma$, E$_{cut}$ and reflection.

The results of the fitting procedure with model A, B and C  are presented in Table \ref{modelA}, Table \ref{modelB} and Table \ref{modelC} respectively.

\begin{sidewaystable*}
\begin{flushleft}
\caption{\bf{Model A} composed by an absorbed power-law
  component, a thermal soft emission and a gaussian iron line.}
\begin{tabular}{lccccccccccc}
\noalign{\hrule}
\noalign{\medskip}
& \textbf{$^1$N$^{\rm gal}_{\rm H}$ }& \textbf{$^1$N$_{\rm H}$}  & \textbf{$\Gamma$} &
\textbf{A$_{soft}$/A$_{IC}$} & \textbf{$^2$KT$_{soft}$}
& \textbf{$^2$E$_{line}$} & \textbf{$^2\sigma_{\rm line}$} & \textbf{$^3$EW$_{\rm line}$} & \textbf{$^4$F$^{unabs}_{2-10keV}$} & \textbf{$^4$F$_{20-100keV}$}& \textbf{$\chi^2$/dof} \\
 & & & & & & & & & & & \\
\hline
 & & & & & & & & & & & \\
\leda & 0.0369 & 4.4$^{+1.2}_{-1.5}$ & 1.7$^\star$ & 0.01 &
 0.9$^{+0.1}_{-0.1}$  & 6.4$^\star$ & 0.01$^\star$ & $<$ 1000 & 2.7 & 4.9 & 16/24 \\
 & & & & & & & & & & & \\
\hline
 & & & & & & & & & & & \\
\075 & 0.458  & 0.4$^{+0.2}_{-0.2}$ & 1.7$^\star$ & - & - & 6.4$^\star$ & 0.01$^\star$
& $<$ 548  & 0.8 & 1.5 & 68/51 \\
 & & & & & & 2.4$\pm$0.5 & $<$0.2 & 109$^{+88}_{-78}$& & & \\
\hline 
 & & & & & & & & & & & \\
\120 & 0.164 & $31^{+120}_{-27}$ & 1.7$^\star$ & 0.12& $1.7^{+0.1}_{-0.2}$ 
&6.4$^\star$ & 0.01$^\star$  & $<$ 3000 & 10 & 3.9 & 57/50 \\
 & & & & & & & & & & & \\
\hline
 & & & & & & & & & & & \\
\ngc & 0.0217 & $40^{+6}_{-2}$ & $1.6^{+0.1}_{-0.1}$ &  0.01 & $0.30^{+0.10}_{-0.07}$ & 6.3$^{+0.1}_{-0.1}$
& 0.27$^{+0.23}_{-0.15}$& 243$^{+177}_{-80}$ & 1.7 & 3.8  & 304/297 \\
 & & & & & & & & & & & \\
\hline
 & & & & & & & & & & & \\
\eso & 0.0764 & $18.8^{+0.7}_{-0.7}$ & $1.88^{+0.06}_{-0.07}$
& 0.002 &$0.28^{+0.05}_{-0.04}$ &$6.3^{+0.1}_{-0.1}$ & 0.47$^{+0.11}_{-0.11}$
&$197^{+33}_{-33}$ & 6.1 & 7.9 & 1078/1101\\
 & & & & & & & & & & & \\
\hline
 & & & & & & & & & & & \\
\ic & 0.09 & 15.0$^{+1.7}_{-2.7}$ & 1.3$^{+0.1}_{-0.1}$ & 0.03 & 0.22$^{+0.03}_{-0.03}$ &
$6.26^{+0.07}_{-0.14}$ & $<0.3$ & $335^{+197}_{-166}$ & 0.58 & 2.9 & 76/125 \\
 & & & & & & & & & & & \\
\hline
 & & & & & & & & & & & \\
\j104 & 0.137 & 4.1$^{+0.3}_{-0.1}$ & 1.83$^{+0.06}_{-0.05}$ & 0.009 & 0.39$^{+0.08}_{-0.08}$ &
$6.21^{+0.05}_{-0.07}$ & 0.19$^{+0.11}_{-0.06}$ & $209^{+67}_{-50}$ & 1.7 & 2.5 & 1132/1070 \\
 & & & & & & & & & & & \\
\hline
\end{tabular}
\label{modelA}
\end{flushleft}
\small{
$^{(1)}$ In 10$^{22}$ \cmM2.
$^{(2)}$ In keV.
$^{(3)}$ In eV.
$^{(4)}$ In 10$^{-11}$ erg cm$^{-2}$ s$^{-1}$.
$^\star$ Frozen during the fit.}
\end{sidewaystable*}
%

\begin{sidewaystable*}
\begin{flushleft}
\caption{\bf{Model B} composed by an absorbed power-law
  component with high energy exponential cut-off, a thermal soft emission and a gaussian iron line.}
\begin{tabular}{lcccccccc}
\noalign{\hrule}
\noalign{\medskip}
& \textbf{$^1$N$_{\rm H}$}  & \textbf{$\Gamma$ }& \textbf{$^2$E$_c$} & \textbf{$^2$KT$_{soft}$} &
\textbf{$^2$E$_{line}$ }
& \textbf{$^3$EW$_{\rm line}$ }& \textbf{$^2\sigma_{\rm line}$ }& \textbf{$\chi^2$/dof }\\
 & & & & & & & &  \\
\hline
 & & & & & & & & \\
\leda & 4.2$^{+1.5}_{-1.5}$ &1.7$^\star$ & $>$160
&  1.1$^{+2.9}_{-0.3}$ & 6.4$^\star$ & $<$1600 & 0.01$^\star$ & 18/23\\
  & & & & & & & & \\
\hline
 & & & & & & & & \\
\075 & 0.5$^{+0.2}_{-0.3}$ &1.7$^\star$ & $>$80
& - &  6.4$^\star$& $<$540 &  0.01$^\star$ & 58/50\\
  & & & & & 2.4$^{+0.2}_{-0.1}$ & 108$^{+92}_{-71}$& $<$0.2& \\
\hline
 & & & & & & & & \\
\120 & 15$^{+39}_{-13}$ &1.7$^\star$ & $83^{+167}_{-13}$
&1.6$^{+0.3}_{-0.1}$ & 6.4$^\star$ & $<$2000 & 0.01$^\star$&  43/49\\
 & & & & & & & & \\
\hline
 & & & & & & & & \\
\ngc & 33.8$^{+2.8}_{-3.6}$ &1.3$^{+0.3}_{-0.1}$ & $60^{+50}_{-20}$
&0.29$^{+0.10}_{-0.08}$ & 6.3$^{+0.1}_{-0.1}$ & 334$^{+84}_{-103}$
& 0.4$^{+0.2}_{-0.1}$ & 297/296\\
 & & & & & & & & \\
\hline 
 & & & & & & & & \\
\eso & 18.6$^{+0.8}_{-0.8}$ &1.82$^{+0.09}_{-0.15}$ & $>$70
&0.28$^{+0.04}_{-0.04}$ & 6.28$^{+0.11}_{-0.10}$ & 189$^{+63}_{-50}$
& 0.47$^{+0.13}_{-0.11}$ & 1076/1100\\
  & & & & & & & & \\
\hline 
 & & & & & & & & \\
\ic & 15$^{+3}_{-2}$ &1.2$^{+0.1}_{-0.2}$ & 133$^{+280}_{-75}$
&0.22$^{+0.03}_{-0.03}$ & 6.27$^{+0.07}_{-0.13}$ & 297$^{+184}_{-141}$
& $<1.0$ & 74/124\\
 & & & & & & & &\\
\hline
 & & & & & & & & \\
\j104 & 4.10$^{+0.03}_{-0.07}$ &1.82$^{+0.04}_{-0.06}$ & $>$170 & 
0.39$^{+0.10}_{-0.07}$ & 6.21$^{+0.05}_{-0.07}$ & 210$^{+65}_{-48}$
& $0.19^{+0.11}_{-0.04}$ & 1132/1069\\
 & & & & & & & &\\
\hline
\end{tabular}
\label{modelB}
\end{flushleft}
\small{
$^{(1)}$ In 10$^{22}$ \cmM2.
$^{(2)}$ In keV.
$^{(3)}$ In eV.
$^{(4)}$ In 10$^{-11}$ erg cm$^{-2}$ s$^{-1}$.
$^\star$ Frozen during the fit.}
\end{sidewaystable*}
%

\begin{sidewaystable*}
\begin{flushleft}
\caption{\bf{Model C} composed by an absorbed
  e-folded power-law component, a reflection from a cold slab, a thermal soft emission and a gaussian iron line.}
\begin{tabular}{lccccccccc}
\noalign{\hrule}
\noalign{\medskip}
 & \textbf{$^1$N$_{\rm H}$}  & \textbf{$\Gamma$ }& \textbf{$^2$E$_C$} &  \textbf{$^\dagger$R} & \textbf{$^2$KT$_{soft}$}
& \textbf{$^2$E$_{Line}$} & \textbf{$^3$EW$_{\rm line}$ }& \textbf{$^2\sigma_{line}$} & \textbf{$\chi^2$/dof} \\
& & & & & & & & &  \\
\hline
& & & & & & & & &\\ 
\leda & 2.9$^{+1.3}_{-0.3}$ & 1.7$^\star$ & $>$210  & $<$2.4($<$2.4) &
1.3$^{+1.7}_{-0.3}$ &6.4$^\star$ & $<$1800 & 0.01$^\star$ &18/22 \\
& & & & & & & & & \\
\hline
& & & & & & & & & \\
\075 &0.4$^{+0.2}_{-0.3}$ & 1.7$^\star$ & 40$^{+160}_{-10}$ & $<$7 ($<$7) &
- & 6.4$^\star$ & $<$621 & 0.01$^\star$& 55/49\\
& & & & & &  2.46$^{+0.08}_{-0.05}$ &184$^{+87}_{-76}$ & $<$0.2& \\
\hline
& & & & & & & & & \\
\120 & 3.3$^{+41}_{-2.7}$& 1.7$^\star$ &  170$^{+130}_{-50}$ & 1.3$^{+5.7}_{-0.4}$ (1.3$^{+6.7}_{-1.2}$) &  1.8$^{+0.3}_{-0.3}$ & 6.4$^\star$  & $<$3000 & 0.01$^\star$ & 40/48\\
& & & & & & & & &  \\
\hline
& & & & & & & & & \\
\ngc & 30$^{+6}_{-3}$ &  1.25$^{+0.05}_{-0.17}$ &
62$^{+38}_{-24}$ &  0.9$^{+1.3}_{-0.7}$ ($<$2.5)&  0.29$^{+0.11}_{-0.06}$ &
6.31$^{+0.11}_{-0.13}$ &  539$^{+231}_{-109}$ &0.44$^{+0.28}_{-0.29}$ & 296/295 \\
& & & & & & & & & \\
\hline
& & & & & & & & & \\
\eso & 18.9$^{+0.6}_{-1.1}$ &  1.9$^{+0.2}_{-0.2}$ &
50$^{+250}_{-25}$ &  1.7$^{+0.6}_{-0.5}$ (2.1$^{+1.5}_{-1.1}$)&  0.28$^{+0.04}_{-0.03}$ &
6.27$^{+0.11}_{-0.12}$ &  325$^{+100}_{-100}$ &$<$0.06 & 1062/1099 \\
& & & & & & & & & \\
\hline
& & & & & & & & &\\
\ic & 14$^{+3}_{-1}$ &  1.5$^{+0.2}_{-0.3}$ &
70$^{+60}_{-30}$ &  2.6$^{+3.5}_{-1.5}$ (1.9$^{+1.3}_{-1.3}$) &  0.23$^{+0.03}_{-0.02}$ &
6.28$^{+0.05}_{-0.05}$ &  360$^{+106}_{-190}$ & $<$ 0.35 & 67/123 \\
& & & & & & & & & \\
\hline
 & & & & & & &  \\
\j104 & 4.6$^{+0.3}_{-0.4}$ &  1.95$^{+0.28}_{-0.07}$ &
$>$120 & 1.1$^{+2.4}_{-0.7}$ (2.2$^{+3.2}_{-1.5}$) & 0.39$^{+0.20}_{-0.05}$ &
6.22$^{+0.03}_{-0.06}$ &  215$^{+70}_{-80}$ & 0.13$^{+0.12}_{-0.06}$  & 1120/1068 \\
 & & & & & & &\\
\hline
\end{tabular}
\label{modelC}
\end{flushleft}
\small{
$^{(1)}$ In 10$^{22}$ \cmM2.
$^{(2)}$ In keV.
$^{(3)}$ In eV.
$^{(4)}$ In 10$^{-11}$ erg cm$^{-2}$ s$^{-1}$.
$^\star$ Frozen during the fit.
$^\dagger$ Reflection fraction obtained with C$_{INTEGRAL}$=1 and C$_{INTEGRAL}$ free to vary (in the parentheses).}
\end{sidewaystable*}
%

\section{Results and Discussion}
\label{RD}
All our models provide a good fit to all spectra, with $\chi^2$/dof around
1. Model A, B and C have in fact an average reduced $\chi^{2}$ value of 0.97, 0.92 and 0.90 respectively. Here we present the main results of our analysis.

\subsection{The soft X-ray spectra}
All sources with the exception of \075 show evidence for an unabsorbed component at energies less
than 2 keV.
Recent high resolution spectroscopic data available with \xmm\ on bright absorbed AGN (Massaro et al. 2006, \cite{cielo}) show evidence that the soft excess component is dominated by emission lines and originates in a gas photoionized by the primary continuum in which the density is decreasing with radius like $r^{-2}$, as observed in the Narrow Line Region (NLR). On the other hand high resolution images available with \chandra\ show that in some cases (e.g. NGC 4549 Schurch et al. 2002) the soft X-ray emission has a thermal origin and is dominated by a starburst component. 
We have therefore taken two approaches, in the first we modelled the soft excess with a thermal component, i.e. a black-body model.
With only one exception, the temperatures we found are in the range 0.2--0.9 keV (see Table \ref{modelA}); in the case of \120\ the value of kT=1.7$^{+0.1}_{-0.2}$ keV is much higher than typically observed in type 2 AGN. In the second, we investigated the  alternative scenario  in which the soft X-ray excess is reproduced by a
scattered component, i.e.  a power-law having the same photon index of the nuclear continuum but a different normalization. 
In this case  we still obtain a good fit
in the case of \leda, \ngc, \eso, \ic\  and \j104 ($\chi^2$/dof are 25/25, 291/298, 1081/1102, 102/126 and 
1136/1071 respectively to be compared
with those in Table \ref{modelA}). The only exception is again \120. 
The value of the ratio A$_{IC}$/A$_{soft}$  (relative to the case where the soft and hard photon indices are frozen to the same value) for all sources  is listed 
in Table \ref{modelA}. This value provides an estimate of the  relative strength of the soft excess with
respect to the primary emission. The ratios in Table \ref{modelA}  are of the
order of few per cent for those sources that are well reproduced by
a soft power-law component; this value is also in good agreement with that measured in a sample of absorbed AGN observed with 
\xmm\ and \chandra\ (Bianchi et al. 2006).
 This soft component is likely produced
by electron scattering of the intrinsic continuum and is observed in most 
Seyfert 2 galaxies (Antonucci 1993, Matt et al. 1997). 
The evidence that the $\Gamma_{soft}$ and $\Gamma_{hard}$ are
consistent with each other and that A$_{IC}$/A$_{soft}$ is in the expected range of values, strongly support this scenario. 

In the case of \120 a scattered component is not able to reproduce the soft X-ray excess ($\chi^2$/dof=119/51 when 
we substitute the thermal component with a power-law). Fitting with the scattered model \chandra\ data only, we get a good fit, however the ratio
between the normalization of the two power-laws is $\sim$ 0.12, making this scenario unlikely. The high value of kT and A$_{IC}$/A$_{soft}$ in \120 give indication that the contribution of the soft excess component to the emission is higher than in the other sources. This could be due to a starburst contribution. 
We will come back to this source in the appendix.

If left free to vary, the photon index of the
soft component remains consistent with that of the primary emission in the case of \leda, \eso\ and \ngc\ ($\Gamma_{soft}=1.0^{+0.9}_{-0.7}$, $\Gamma_{soft}=1.8^{+0.4}_{-0.2}$ 
and  $\Gamma_{soft}=2.1^{+0.5}_{-0.5}$ respectively to be compared with the values in Table \ref{modelA}).  This supports the 
''scattered'' scenario for these three sources.
On the contrary for \ic\ if left free to vary the photon index of the soft unabsorbed component is steeper with respect to that of the absorbed one
($\Gamma_{soft}=2.7^{+0.4}_{-0.4}$) suggesting that in this case the soft X-ray component could be of thermal origin.
In the case of \j104 the soft photon index gets instead harder than that of the nuclear component ($\Gamma_{soft}=0.9^{+0.2}_{-0.1}$);
such a hard value is probably related to the presence of the Compton hump above 10 keV. This clearly shows the importance of fitting
simultaneously the data from 0.5 to 200 keV.

\075 is a peculiar case, as in this source we do not find evidence for  "continuum'' soft X-ray emission. 
A narrow emission line at $2.46^{+0.08}_{-0.05}$ keV is required in the model for this source; the line has an equivalent width of 
$ 184^{+87}_{-76}$ eV with respect to the unabsorbed continuum (see Table \ref{modelC}; model C represents the best fit in the case of \075). The energy of the line suggests that it could be due to K$_{\alpha}$ fluorescence of highly ionized elements like S or Si (SXV with the theoretical value of energy of 2.430 keV and SiXIV with the theoretical value of energy of 2.461 keV) and could be produced in a scattering warm  gas photoionized by the primary emission as already observed in other obscured AGN such as NGC 1068 and Circinus galaxy (Guainazzi et al. 1999, Massaro et al. 2006).
The lack of any other emission feature due to high ionized elements prevents the creation of a ionization structure for this gas. 
If we assume that the warm medium has N$_{H} \sim 10^{22}$ \cmM2 and ionization parameter U$_{X}$ (defined as the ratio between the density of ionizing photons and the gas density) in the range  in 0.1--5, we can produce the observed line.  This gas could be associated to the NLR and/or the $\left[ \rm O III\right] $ ionization cones as demonstrated to be the case for other obscured AGN (\cite{bianchi06}).

\subsection{Absorption}

A column density in excess to the Galactic value  was measured in all sources, as expected in type 2 Seyfert 
galaxies. The minimum and maximum values of N$_H$ in our sample are
0.1 and 40$\times 10^{22}$\cmM2 respectively, with an average value
$\bar{N}_H$=(10.6$^{+0.2}_{-0.2})\times 10^{22}$ \cmM2 (see Table \ref{modelC}). 
For all our sources we have also estimated the F$_X$/F[OIII]$\lambda$5007 ratios (see Table \ref{OIII}). 
The flux of [OIII] $\lambda$5007 has been corrected for intrinsic galactic reddening following the prescription of Bassani et al. (1999) 
 and is used here as an isotropic indicator of the intrinsic brightness of each source. We find that all F$_X$/[OIII] ratios are indeed well above those measured in Compton thick sources and comparable with those observed in type 1 AGN (Bassani et al. 1999).
The column densities measured as well as the F$_X$/F[OIII]$\lambda$5007 ratios, shown in Table \ref{OIII}, allow us to confirm the already established Compton thin nature in the case of \leda, \075, \120 and \eso\ and to establish that also \ngc, \ic\ and \j104\ are Compton thin AGN.

\begin{table}
\caption{Fux of [OIII] $\lambda$5007  corrected for absorption and its ratio with the unabsorbed flux in 2--10 keV.}
\begin{tabular}{lccc}
\hline
source & $^\star$F([OIII]) & F$_X$/F[OIII] & ref \\
\hline
\leda & 0.55 & 4.9 & (1)\\
\075 & $<$ 0.02 & $>$ 40 & (2) \\
\120 & 0.2454 & 40.7 & (2)\\
\ngc & 0.027 & 62.9 & (3)\\
\eso & 0.112 & 54.5& (4) \\
\ic & - & - &  -\\
\j104 & 0.004 & 425 & (5) \\
\hline
\end{tabular}
\label{OIII}

\small{$^\star$ In 10$^{-11}$ erg cm$^{-2}$ s$^{-1}$. (1) \cite{masettiII}. (2) \cite{masettiV}. (3) Vaceli et al. 1997. (4) Bassani et al. 1999. (5) \cite{masettiIII}}
\end{table}

Malizia et al. (2007) recently analysed a sample of AGN observed with \integral\ and \swift\ providing a diagnostic tool (Flux (2-10 keV)/F(20-100 keV) softness ratio) to isolate peculiar objects  and to investigate their Compton thin/thick nature. 
Using this new tool they were able to confirm again the Compton thin nature of all our sources, but identified \075 as a peculiar object as its location in this new diagnostic plot was outside the expected trend.

\subsection{The reflecting medium, the origin of the Compton
bump and of the iron line}

\begin{figure}
\centering
\includegraphics[width=0.7\linewidth]{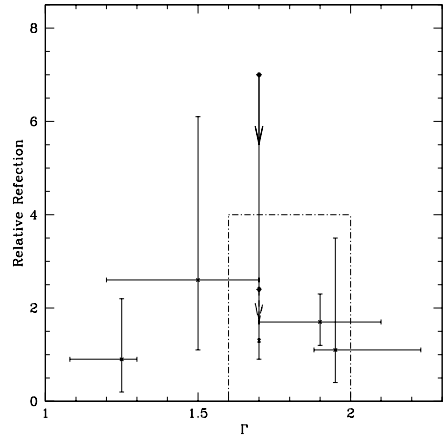}
\caption{Reflection fraction versus photon index in our sample. Dot-dash box show the average values found in the \sax\ sample of type 2 AGN  (\cite{risaliti02}).}
\label{gamma_refl}
\end{figure}

We measure the reflection fraction through the parameter R that
represents the ratio between the reflected and primary
components. The model (PEXRAV, \cite{pexrav_ref}) assumes a perfect reflection from an
infinite slab covering 2$\pi$ angle. This means that in the case of a
face-on line of sight, R has to be in the range 0--2. 
The importance of the reflection component is given by a comparison
between Table \ref{modelB} and Table \ref{modelC}.
The addition of one interesting parameter, R in model C, with respect to model B is statistically
required in the case of \eso, \ic\ and \j104, with a significance of
 99.98, 99.95 and 99.93 per cent
respectively, according to the standard F-test. In the case of \120 and \ngc, R is constrained but with a F-test probability of only 93.6 and 99.1 per cent respectively.
In the case of \leda\ and \075 we are only able to find an upper limit to R 
even if with an unphysical value greater than 2. We will return to this point
later. 
In Figure \ref{gamma_refl} we plot the reflection fraction R versus the
photon index $\Gamma$. The values of R we obtain are in good agreement with those found in the \sax\ sample of type 2 AGN  (\cite{risaliti02}), \sax\ average values are also shown in Figure \ref{gamma_refl}. \ic\ and \ngc\ show evidence of a flatter photon index than the other sources, this effect will be discussed in Sect. \ref{continuum} and Appendix \ref{single sources}. No evidence for the correlation claimed by Zdziarski
et al (1999) between R and $\Gamma$ is found (correlation coefficient $r = 0.0582$, suggesting no real evidences against the null hypothesis).
It is important to stress here that the value of the Compton
reflection fraction R is strongly dependent on the value of the cross-calibration constant C$_{INTEGRAL}$ between the soft gamma 
(\integral) and X-ray data (\chandra, \xmm\ and \asca), that we assumed to be equal
to one in all spectra.
It is important to note that Kirsch  and collaborators (2005) analysed the Crab spectrum observed with different instruments, and they concluded that at 20 keV the cross-calibration \xmm/\integral\, \chandra/\integral\ and \asca/\integral\  was close to 1 (within fews per cent). This strengthens  the assumption we made on C$_{INTEGRAL}$. However we also checked our assumption \textit{a posteriori}. In the case of \xmm\ data, if left free to vary in the model C, the cross-calibration constant MOS1/ISGRI ranges between 0.5 and 2.
In the case of \asca\ data of \ngc, C$_{INTEGRAL}$ is 0.8$^{+0.6}_{-0.6}$. In the case of \chandra\ data, C$_{INTEGRAL}$ was much less constrained with respect to \xmm\ and \asca, mainly because the larger gap between IBIS and ACIS (7--20 keV). In fact for \chandra\ data, if left free to vary in model C, the cross-calibration constant is found in the range 0.5-4; however the photon index fixed to 1.7 in the fit, still allow us to measure the reflection fraction R in \120, while only upper limits are found in \leda\ and \075.
In table \ref{modelC} we report the value of R we find leaving C$_{INTEGRAL}$ free to vary, and it is evident that, even if the value of R are less constrained, we are able to measure the Compton Reflection in four objects, and their best fit value are not very different from those obtained with C$_{INTEGRAL}$=1.
We stress also that the errors associated to R are higher than
those of $\Gamma$ or kT and this is due to the degeneracy between the
different spectral parameters, in particular between $\Gamma$, E$_c$
and R. If we keep fixed $\Gamma$ and/or E$_c$, the values of R are much
more constrained with respect to those reported in Table \ref{modelC} and
Fig. \ref{gamma_refl}. \\
The simplest scenario to explain the presence of the Compton hump
above 10 keV in type 2 AGN, is one 
in which the reflecting and absorbing medium are the same, probably the putative molecular torus.
Within this scenario the question to address is if the reflection fraction we find is consistent with the
absorbing column densities we measured.
For N$_H=10^{23}-10^{24}-10^{25}$ \cmM2, the contribution of the torus to the
30 keV flux is 8, 29 and 55 per cent respectively (Ghisellini et
al. 1994). This means that the value of R measured in our sources is too high to be associated only
to the absorbing gas of the torus (see Table \ref{modelC}). A possible solution to this problem is that the absorber
is not homogeneous and that a Compton thick medium covers a large
fraction of the solid angle but not the line of sight as proposed by
Risaliti (2002). This absorber can be identified with the clumpy torus recently proposed to explain infrared observations of AGN and their  classification in the unification theory scenario (Elitzur \& Shlosman 2006).

\begin{figure*}
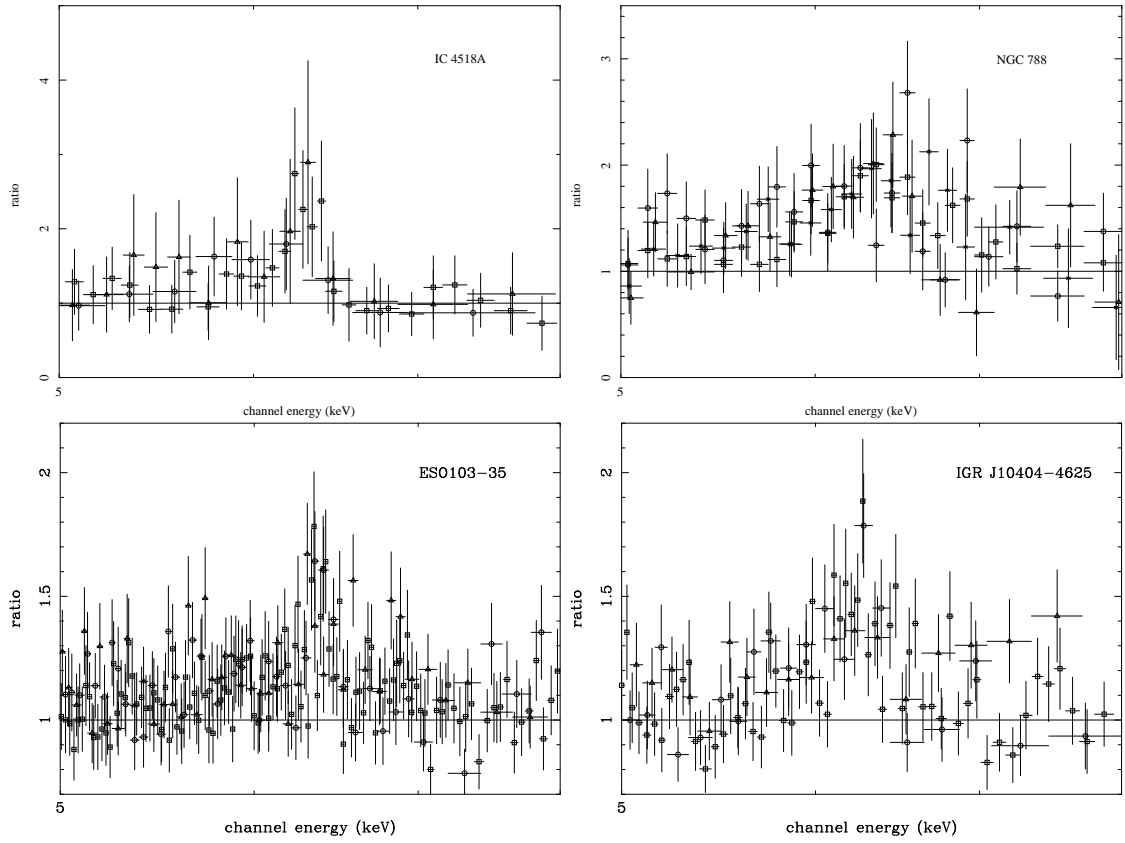

\centering
\includegraphics[width=0.3\linewidth,angle=-90]{fe_ic4518.ps}
\includegraphics[width=0.3\linewidth,angle=-90]{fe_788.ps}
\includegraphics[width=0.3\linewidth,angle=-90]{fe_eso.ps}
\includegraphics[width=0.3\linewidth,angle=-90]{fe_igr104.ps}
\caption{The iron line profile when the continuum is reproduced
  with model C. For \ngc\ stars, triangles, circles and squares
  represents SIS0, SIS1, GIS2 and GIS3 respectively For \ic,
  \eso\ and \j104 triangles, circles and squares represent MOS1, MOS2
  and pn data respectively.}
\label{fe_line}
\end{figure*}

\begin{figure*}
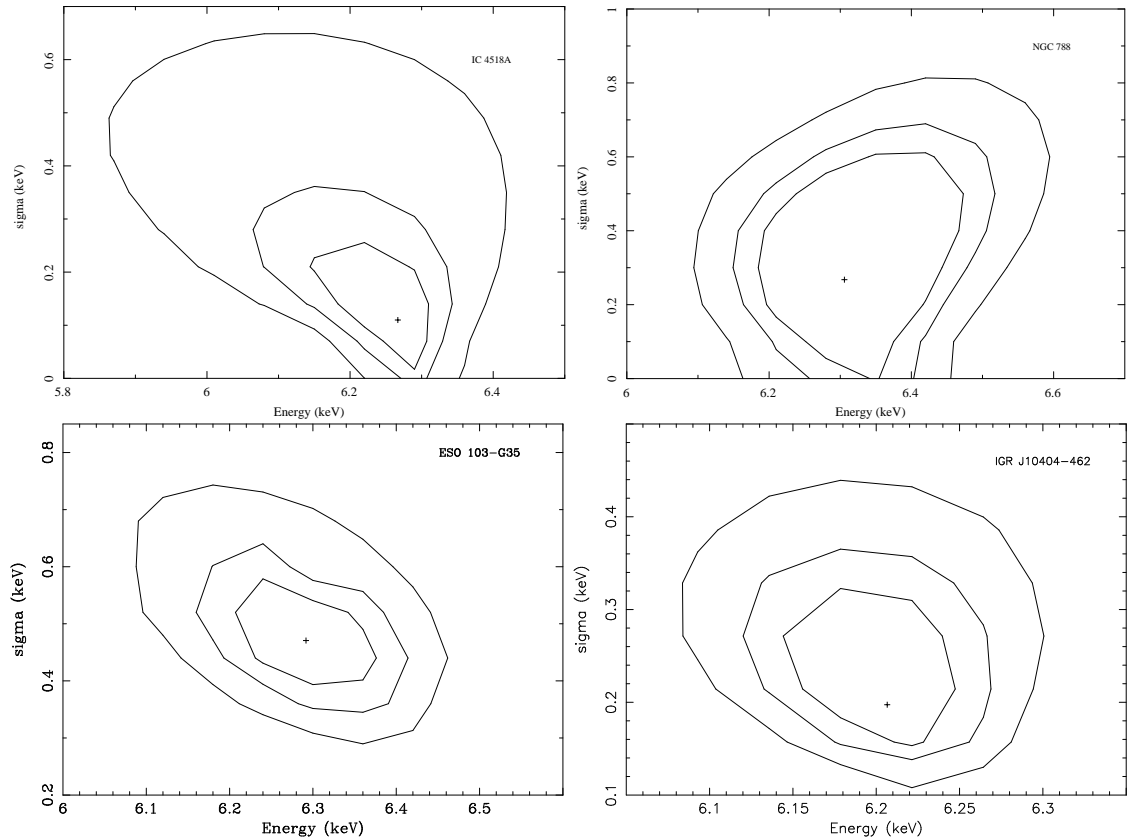

\centering
\includegraphics[width=0.3\linewidth,angle=-90]{ene_sigma_fe_ic.ps}
\includegraphics[width=0.3\linewidth,angle=-90]{ene_sigma_fe_ngc.ps}
\includegraphics[width=0.3\linewidth,angle=-90]{ene_sigma_fe_eso.ps}
\includegraphics[width=0.3\linewidth,angle=-90]{ene_sigma_fe_104.ps}
\caption{Contour plots 99, 90, 68 per cent of the intrinsic width $\sigma$ vs Energy for the iron lines.}
\label{cont_fe_line}
\end{figure*}

In the case of \ngc, \eso, \ic\ and \j104, it has been possible to measure the K$\alpha$ iron
line. For the threee \chandra\ sources, \leda, \075\ and \120, we have freozen the energy of the (narrow) line E$_{Fe}$=6.4 keV  and we were able to find only upper limits on the equivalent width. 
For this reason we present our conclusions about Fe line on the basis of the results we found in the other four sources. 
The profiles of the line when the continuum is fitted with
model C, are shown in Figure \ref{fe_line}. 
The energy of the line for all the objects is consistent with 
neutral or little ionized iron. The line is narrow in all objects as shown by
the contour plot of $\sigma$ vs Energy (see Fig. \ref{cont_fe_line}) obtained fitting data with Model A.  
In the case of \eso\ the contours plot are only marginally consistent with a narrow line profile ($\sigma$=0.47$^{+0.11}_{-0.11}$ keV with Model A, see Table \ref{modelA}), however when the continuum is well reproduced with Model C, in which all the spectral components are modelled, we find an upper limit to the intrinsic width of the line ($\sigma <$ 0.06 keV, see Table \ref{modelC} a Figure \ref{fe_line}) consistent with a narrow profile, as already observed by \sax\ (Akylas et al. 2001). This underlines the importance of a good modeling of the continuum to investigate the line properties.

\begin{figure}[t]
\centering
\includegraphics[width=0.7\linewidth]{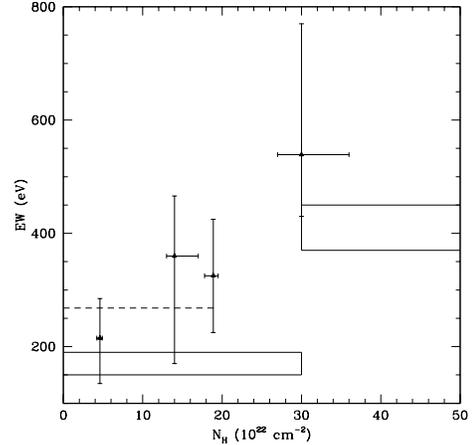}
\caption{Absorbing column density versus Fe equivalent width. The
  dotted line represent the best fit constant in the region
  N$_H<$3$\times$10$^{23}$ \cmM2. The box regions represent the average values of N$_H$ (1$\sigma$ errors) found in the \sax\ sample of type 2 AGN  (\cite{risaliti02}).}
\label{nh_ew}
\end{figure}

In Figure \ref{nh_ew} we plot the equivalent width versus the column
density for the four sources where the Fe line has been measured. If the Fe line is produced far away form the central source, a higher value of the column density will absorb a higher fraction of the continuum at the energy of the line, but not the line photons; the total effect will be a larger value of the equivalent width.
The errors on EW are too large to draw solid conclusion about a possible
correlation, however fitting the EW data shown in Figure \ref{nh_ew}  with a constant K we found a value K=285$\pm$57 eV,
with $\chi^2$/dof=1.2. The same fit in the region below
3$\times$10$^{23}$ \cmM2 gives K=268$\pm$60 eV, with $\chi^2$/dof=1 (this best fit constant it plotted in Figure \ref{nh_ew} with a dashed line).
The only data point with N$_{H} >3 \times$10$^{23}$ \cmM2 is that of \ngc\ which has a  N$_{H}$ significantly higher than the other three sources.
Even if we do not have enough statistics to draw any firm conclusions, 
we stress that the trend we find confirms the results of Risaliti (2002) on a sample of Seyfert 2 observed by \sax. \sax\ N$_{H}$ average values below and above 3 $\times$10$^{23}$ \cmM2 are also plotted in Figure \ref{nh_ew}.
This behaviour represents strong evidence that
 the line has to be produced (at least in part) far away from the accretion disc, being absorbed in a different way 
than the continuum. The evidence of a narrow profile (see Fig. \ref{cont_fe_line}) strongly supports this scenario,
 and confirms that the best candidate for the line production site is the absorbing gas, probably a non homogeneous obscuring torus.

The question is now if the absorbing gas alone, the torus, is able to produce
the observed iron lines.  The value of N$_H$ in the four sources where we measure the iron line is
in the range $(4.2-36)\times 10^{22}$\cmM2 with an average value
(16.9$^{+0.2}_{-0.2})\times 10^{22}$ \cmM2 (see Table \ref{modelC}). 
A torus with column densities in this range could produce a fluorescence iron line with an average equivalent 
width in the range 10-200 eV  and with an average value of $\sim$100 eV (Ghisellini et al. 1994).
These values are well below those we measure suggesting that the line has to be produced, at least partly, in a gas that is different from the absorbing one. Another possibility is, again, a non homogeneous absorber not covering the line of sight uniformly. The upper limits of the equivalent width  we found in the three \chandra\ spectra also support this conclusion.

The measured values of the reflection components in our sample (both iron line equivalent width and reflection continuum) 
suggest that the absorption is more effective than the
reflection, e.g. making the hypothesis that the absorbing/reflecting
medium is not uniform, likely a clumpy torus (Elitzur \& Shlosman 2006). Another scenario could be that the source is observed throughout a torus with a very shallow angle, in this case the flux directly seen by the observer is due to the inverse Compton
interaction of the hot plasma electrons off the seeds photons.
This high energy radiation exiting the system with a smaller angle with respect
to the torus plan, will interact with a sort of "grazing incidence" with
the upper border of the torus and will see a very high column density
(almost infinite). Conversely, the Compton scattered/reflected photons
escaping in the direction of the line of sight, will have a Compton cone
with a moderate optical path in our lign of sight. This do not apply to
the lower energies, fully absorbed anyway.

\subsection{The intrinsic continuum and the high energy cut-off}
\label{continuum} 
We measure the high energy cut-off in five out of seven sources in the sample (see Table \ref{modelC}), and we
find a lower limit to E$_{c}$ in the case of \leda\ and \j104.
All these values (either measured or lower limits) are within the range found in other Seyfert galaxies (Perola et al 2000).

\begin{figure}
\centering
\includegraphics[width=0.7\linewidth]{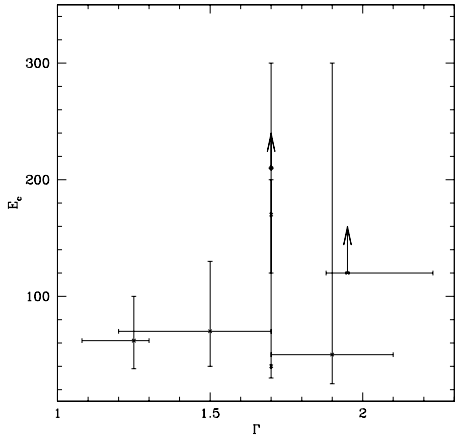}
\caption{High energy cut--off in keV versus photon index}
\label{gamma_ec}
\end{figure}

The phenomenological model we employ to fit the intrinsic continuum (i.e.
power-law with a high energy cut-off), provides a good description of a
 two-phase model involving a hot corona emitting medium-hard X-rays by Comptonization and a cold
optically thick layer (the disk) that provides the soft
photons for the  Comptonization (Haardt \& Maraschi 1991;
Haardt, Maraschi \& Ghisellini 1997). 
In Fig. \ref{gamma_ec} we plot the
high energy cut-off $vs$ the photon index. It is important to note that the value of E$_c$ to be considered are those
listed in Table \ref{modelC}, because above 10 keV the presence of Compton reflection can modify the continuum slope and, in turn,  
the value of the high energy cut-off (we stressed  in the previous sections that there is a strong degeneracy between the reflection fraction R and E$_c$). Fig. \ref{gamma_ec} shows that all the values we constrain are below 300 keV, strongly suggesting the presence of a high energy cut-off in the spectra of Seyfert 2 galaxies.

In \ngc\ and \ic\ we find a value of the photon index lower with respect the average observed in Seyfert galaxies.
The flat photon index can be either due a bad modeling of the primary continuum, and this will be discussed for the two sources in Appendix \ref{single sources}, or it could be real, as expected in the X-ray background synthesis model (\cite{gilli07}). The latter hypothesis needs more statistics to be confirmed.

\section{Conclusions}
\label{conclusions}

We have presented spectral broad-band analysis of a small sample of seven absorbed Seyfert 2 selected at E$>$ 10 keV by \integral/IBIS.  The combined 0.2--200 keV spectra allowed us to perform a detailed study of the emission/absorption properties and to test possible correlations between different parameters (e.g. R vs $\Gamma$ or $E_{c}$ vs $\Gamma$); in particular, for the first time in these sources a measurement of the reflection components (both Compton bump and iron line) and high energy cut-off has been performed.
An \textit{a posteriori}
check of our analysis procedure, which is affected by the limitation of  
using non simultaneous X-ray soft gamma-ray data, is performed providing 
for one well known source, \eso, a direct comparison between our study
and previous ones with \sax\ data. The values of $\Gamma$,
R and N$_{H}$ we found for this source are completely consistent with those
obtained by Akylas et al. (2001) analysing simultaneous broad--band \sax\ observations. This evidence strengthens the validity of our fitting procedure and makes  reliable our results.
The main conclusions or our analysis are the followings.

\begin{itemize}
\item The average value of the absorbing column density is $\bar{N}_H$ = (10.6$^{+0.2}_{-0.2}$) $\times$ 10$^{22}$ \cmM2, and the range observed suggests a Compton thin nature for all objects in the sample, that is a new results in the case of \ngc, \ic\ and \j104. This evidence is also confirmed by the ratio F$_x$/F$\left [ \rm OIII\right ]$ $\lambda$5007.

\item The continuum was reproduced with an e-folded power-law having a photon index in the range 1.3-1.9 and a high energy cut-offs well below 300 keV; this finding suggests that a cut-off is a common factor in Seyfert 2 galaxies with values in good agreement with those found in type 1 AGN.

\item We measure the Compton reflection component in five sources (\120, \ngc, \eso, \ic\ and \j104) and found an upper limit in two others (\leda\ and \075). No evidence for a correlation between R and $\Gamma$ is found. We also observed the iron line (in the case of \chandra\ spectra we found upper limits only on the equivalent width). Both reflection components are not immediately consistent with production in the cold absorbing gas identified in the molecular torus. Both the reflection fraction R and the equivalent width of the line are too high to be produced in the gas with the observed column density. A possible solution is that the absorption is more effective than the reflection, e.g. making the hypothesis that the absorbing/reflecting medium is not uniform, like a clumpy torus, or that the source is observed through a torus with a very shallow opening angle.

\item A soft excess component emerging below 2--3 keV is found in all sources with the only exception of \075.
This component is well reproduced by a thermal black body model having temperature in the range 0.2-0.9 keV, as typically observed in Sy 1.
Alternatively a power-law with photon index equal to that of the illuminating continuum  is also able to reproduce this excess is some cases; the ratio of soft versus primary continuum is of the order of a few per cent, as typically observed in Sy 2. A different case is that of \120 which has, in the thermal model, a temperature kT=1.7$^{+0.1}_{-0.2}$ keV and, in the scattered scenario, A$_{IC}$/A$_{soft}$=0.12; both value are higher than observed in the other sources and, typically, in Seyfert. This strongly suggests that the source is characterized by a larger soft excess (possibly due to a starburst contribution) that is not present in the other sources.

\item \075 is the only source in the sample that does not show any evidence for continuum excess at low energies.
A narrow emission line around 2.5 keV is instead found in the \chandra\ data.

\end{itemize}

We acknowledge the Italian Space Agency financial and programmatic support via contracts
I/R/046/04 and I/023/05/0. We thank the referee for constructive suggestions.

\section*{References}
\frenchspacing
\small
\begin{description}
\bibitem[]{}Akylas, A.; Georgantopoulos, I.; Comastri, A. 2001, MNRAS, 324, 521
\bibitem[Anders \& Grevesse 1989]{}Anders E. \& Grevesse N. 1989, GeCoA, 53, 197
\bibitem[Antonucci 1993]{antonucci}Antonucci R.R. 1993, ARA\&A, 31, 473
\bibitem[Bassani et al. 1999]{bassani99} Bassani, L., Dadina, M., Maiolino, R. et al., 1999, ApJS, 121, 473
\bibitem[Bassani et al. 2006a]{bassani06a} Bassani L., Molina, M. Malizia A.et al. 2006, ApJ, 636, L65
\bibitem[Bassani et al. 2006b]{bassani06b}Bassani, L.; Malizia, A.; Stephen, J. B.; for the INTEGRAL AGN survey team, 6th INTEGRAL Workshop "The Obscured Universe", Moscow, 2-8 July 2006, arXiv:astro-ph/0610455
\bibitem[Beckmann et al. 2006]{beckmann06}Beckmann, V.; Gehrels, N.; Shrader, C. R.; Soldi, S. 2006, ApJ, 638, 642
\bibitem[Bianchi et al. 2006]{bianchi06} Bianchi, S.; Guainazzi, M. \& Chiaberge, M. 2006, A\&A, 44, 499 
\bibitem[Bird et al. 2006]{bird06} Bird A.J., Barlow E.J., Bassani L. et al. 2006, ApJ 636, 765
\bibitem[Bird et al. 2007]{bird07}Bird, A. J., Malizia, A., Bazzano, A., et al. 2007, ApJS, 170, 175
\bibitem[Corbet et al. 2002]{corbet02} Corbet E. A., Norris R. P:, Heisler C. A., et al 2002, ApJ, 564, 650
\bibitem[]{} Davis J. E. 2001, ApJ, 562, 575
\bibitem[]{} Di Cocco G., Caroli E., Malizia A. et al. 2003, A\&A 411, 189
\bibitem[]{} Elitzur, M. \& Shlosman, I. 2006, ApJ, 648, L101
\bibitem[]{}Ghisellini, G., Haardt, F., Matt, G. 2004, MNRAS, 267, 743
\bibitem[Gilli et al. 2007]{gilli07}Gilli, R., Comastri A., Hasinger, G. 2007 A\&A 463,79
\bibitem[Guainazzi \& Bianchi 2007]{cielo} Guainazzi M. \& Bianchi S., 2007, MNRAS, 374, 1290
\bibitem[Guainazzi et al. 1999]{guaina99}Guainazzi, M.; Matt, G.; Antonelli, L. A. 1999, MNRAS, 310, 10
\bibitem[Haardt \& Maraschi 1991]{HM91} Haardt F. \& Maraschi L. 1991, ApJ, 380, L51
\bibitem[Haardt, Maraschi \& Ghisellini 1997]{HMG97} Haardt F., Maraschi L. \& Ghisellini G. 1997, ApJ, 476, 620
\bibitem[]{}Kirsch, M. G.; Briel, U. G.; Burrows, D.; et l. 2005, SPIE, 5898, 22
\bibitem[]{}Lebrun F., Leray J.P., Lavocat P. et al. 2003 A\&A, 411, 141
\bibitem[]{}Malizia A., Landi R., Bassani, L: et al. 2007, ApJ, 668, 81
\bibitem[Masetti \etal 2006a]{masettiII} Masetti, N.; Mason, E.; Bassani, L.; et al., 2006, A\&A, 448, 547
\bibitem[Masetti \etal 2006b]{masettiIII} Masetti, N.; Pretorius, M. L.; Palazzi, E.; et al., 2006, A\&A, 449, 1139
\bibitem[Masetti \etal 2006c]{masettiV} Masetti, N.; Morelli, L.; Palazzi, E.; et al., 2006, A\&A, 459, 21
\bibitem[Massaro \etal 2006]{massaro06} Massaro E., Bianchi S., Matt G., D'Onofrio E., Nicastro F., 2006, A\&A, 455, 153 
\bibitem[Matt \etal 1997]{MFR96} Matt G., Guainazzi M., Frontera F. et al. 1997, A\&A 325, L13
\bibitem[Magdziarz \& Zdziarski 1995]{pexrav_ref} Magdziarz \& Zdziarski 1995, MNRAS, 273, 837
\bibitem[Perola \etal 2002]{per02}Perola G.C., Matt G., Cappi M., et al. 2002, A\&A, 389, 802
\bibitem[Risaliti 2002]{risaliti02} Risaliti G., 2002, A\&A, 386, 379
\bibitem[Shinozaki et al. 2006]{shino06} Shinozaki, K; Miyaji, T; Ishisaki, Y; Ueda, Y; Ogasaka, Y. 2006, ApJ, 131, 2843
\bibitem[]{} Schurch, N. J., Roberts, T. P., Warwick, R. S. 2002, MNRAS, 335, 241
\bibitem[Sazonov et al. 2005]{sazonov05} Sazonov, S., Churazov, E., Revnivtsev, M., Vikhlinin, A., Sunyaev, R. 2005, A\&A, 444, L37
\bibitem[]{} Ubertini P., Lebrun F., Di Cocco G. et al. 2003, A\&A 411, 131
\bibitem[]{} Vaceli, M. S., Viegas, S. M., Gruenwald, R., de Souza, R. E. 1997, AJ, 114, 1345
\bibitem[]{} Winkler C., Courvoisier T., Di Cocco G. et al. 2003, A\&A, 411, 1 
\end{description}

\begin{appendix}
\section{Analysis of single sources}
\label{single sources}

We describe here the details of the analysis of each individual source, we also compare our results to previously published measurements.\\

\textbf{\leda}  
The fraction of pile-up estimated with PIMMs was
around $>$20 per cent. 
 This absorption column density we found with our broad-band analysis is in good agreement with the result presented in Sazonov et al. (2005). 
In this source the presence of a high energy cut-off and a reflection component is not required (see Table \ref{modelA}, \ref{modelB} and \ref{modelC}). We found a lower limit to E$_{c}$  of $\sim$210 keV and an upper limit to R of $\sim$2.

\textbf{\075}
The fraction of pile-up estimated with PIMMs was
around 36 per cent. The absorption column density found in this source with our broad-band analysis
is fully consistent with the value found by Sazonov et al. (2005)  in their analysis.
Our broad--band analysis shows evidence that this source is peculiar in many ways, as already described in Sect. 4.1.
 We measured the value of the high energy cut-off which is 40$^{+160}_{-10}$ keV, the lowest value of our small sample. This very low value is probably due to the high upper limit to the reflection component $<$7; this high value suggests some different geometry than that used here to model the continuum reprocessed component.

\textbf{\120}
The fraction of pile-up estimated with PIMMs was
around 30 per cent. 
Our analysis provides the absorption column density  completely consistent with
that published by Sazonov et al. (2005).
The enhancement observed in the soft X-ray (either modelled with a thermal component or a scattered component) could suggest the presence of a starburst component.
 Our broad-band analysis well constrains the high energy cut-off E$_{c}$=170$^{+130}_{-50}$ keV, and we measured also a relative reflection R=1.3$^{+5.7}_{-0.4}$.

\textbf{\ngc} 
The photon index  measured with the model C is flatter with respect to the average value observed in Seyfert galaxies, this could be either intrinsic or due to the presence of a high energy cut-off below 100 keV. In fact when fitting the data with model A, we found a photon index $\Gamma$=1.6, fully consistent with that expected in type 2 AGN. 
Another possibility could be that the absorber is more complex than that proposed here, e.g. a partial covering.
We were able to constraint the Compton reflection fraction R=0.9$^{+1.3}_{-0.7}$.

\textbf{\eso}
\integral\ data of \eso\ have been analysed by Molina et. al (2007) jointly with \sax\ data  from an observation performed 
in October 1996.  Their data, however, referred to the 2th \integral\ catalogue (Bird et al. 2006), and had only a net
 exposure of  41 ks. Our results, which are obtained with a longer exposure, are in very good 
agreement with previous results. 
In particular the value they found for the high energy cut-off (68$^{+71}_{-25}$ keV), perfectly matches our value 
(see Table \ref{modelB} and \ref{modelC}). They found an upper limit to the reflection fraction R$<$1.9, again consistent with our 
result (see Table \ref{modelC}).
\xmm\ data of \eso\ have been analysed by \cite{shino06} in 0.5-10 keV energy range. 
With their limited energy band it was not possible to obtain information
 about cut-off and reflection component, however the value of the intrinsic column density and the continuum shape they found is in very good agreement with the results of our  broad-band analysis. We stress that the possibility to cover the energy range above 10 keV allowed us to model all the spectral components and then to extract the intrinsic continuum shape.

\textbf{\ic}
In this object the presence of the reflection component above 10 keV is statistically required by the data. However the best fit value larger than 1 suggests that the geometry of the reflector has to be more complex than that used in this work.
The large contribution of the reflection component above 10 keV is also confirmed by the fact that the fit with model B, without reflection component, gives a very flat photon index.

\textbf{\j104}
The presence of the reflection component above 10 keV is statistically required by the data, with R=1.1$^{+2.4}_{-0.7}$. 
We were able to find a lower limit only to the high energy cut-off, E$_{c}>$120 keV.

\end{appendix}

\end{document}